\documentstyle{ioplppt}
\begin{document}
\title{Possible long time tails in the current-current
correlation function for the two dimensional electron gas in random
magnetic field.}
\author{ Andrzej {\L}usakowski\dag,\ {\L}ukasz A. Turski\ddag}
\address{\dag\ Institute of Physics, Polish Academy of Sciences
 Al.\ Lotnik{{\'o}}w 32/46, 02-668 Warszawa,
 Poland.}
\address{\ddag\  Center for Theoretical Physics, Polish
Academy of Sciences, and College of Science. Al. Lotnik{{\'o}}w
32/46, 02-668 Warszawa, Poland}
\date{\today}
\begin{abstract}
We consider two-dimensional degenerate electron gas in the presence of
perpendicular random magnetic field. The magnetic field disorder which
is assumed to be gaussian is characterized by two parameters. The first
is proportional to the amplitude of local magnetic field fluctuations.
The second one characterizes the disorder on the longer scale - we call
it the screening length. Using Kubo formula for the conductivity we have
found a class of diagrams which leads to the long time tails in the
current - current correlation function. For short times comparing to the
diffusion time corresponding to the screening length this function
behaves like logarithm of time, for longer times it decays like $t^{-1}$
what of course puts in question the diffusive character of the behavior
of the charged particle in random magnetic field widely assumed in the
literature.

\end{abstract}
%
  The theoretical description of two-dimensional, quantum,
noninteracting, degenerate
electron gas moving in a spatially random magnetic field attracts
nowadays considerable attention in view of its possible relation to
theory of Quantum Hall Effect near the filling factor $\nu = 1/2$
proposed by Halperin and collaborators \cite{Halperin}. The
characteristic feature of these systems behavior is their temporal and
spatial dispersion reflecting underlying microscopic motion nonlocality
\cite{Lusak}. Progress in developing general theory capable of
describing this dispersion in degenerate electron gas is rather slow.
 Attempts were directed, therefore, to analysis of relatively
simpler problem, namely the particle diffusion. In the recent work
Aronov, Mirlin and W{\"o}lfle \cite{Aronov} proposed the derivation of the
zero frequency diffusion coefficient for a two-dimensional degenerate
electron gas in random magnetic field. In the present paper we propose
a next step in the direction of the full analysis by using the
Kubo formula for the conductivity. Analyzing a particular class
of diagrams we have found the long time tails in the current
current correlation function.

We consider a gas of noninteracting, charged, spinless fermions
moving on a two-dimensional plane in the presence of a spatially
random  magnetic field ${ \bf B}({\bf  r})$ perpendicular to that plane.
 The single particle  Hamiltonian reads:
\begin{equation}
\label{hamiltonian}
  H=\frac{1}{2m}\left(\frac{\hbar}{i} \nabla-\frac{e}{c}{\bf A}\right)^2
\end{equation}
where ${\bf A}$ is the magnetic field vector potential.

   The paramount difficulty is now the choice of statistical ensemble
for the random magnetic field. The seemingly obvious, and natural,
choice of Gaussian distributed magnetic field with the second moment
$\langle B_z({\bf r})B_z({\bf r}')\rangle \sim   \delta({\bf r}-{\bf r}')$
   results in infrared divergencies in the lowest order terms in the
perturbation theory \cite{Aronov}.  This is due to the long
range vector potential
correlations reflecting the fact that above Gaussian distribution for
magnetic field permits the field configurations in which arbitrary
large magnetic flux is penetrating the plane. The mean square value of
magnetic flux inside a region is proportional to its area, and this
invalidates  perturbational approach to the problem.

   To avoid this we consider a model of the magnetic field disorder in
which the above mentioned configurations are explicitly suppressed.  To
this effect we assume that the components of the vector potential
$A_i({\bf r})$ ($i=x,y$) are the Gaussian distributed quantities with
zero mean, and which second order cumulant, in the Fourier space, is
given as
\begin{equation}
\label{correlation}
 \langle A_i({\bf q})A_j({\bf q}')\rangle=  
(2\pi)^2\delta({\bf q}+{\bf q}')
\left(\frac{\hbar c}{e}\right)^2 \gamma\frac{1}{q^2+\mu^2}
\delta^T_{ij}({\bf q})
\end{equation}
 where $\delta^T_{ij}({\bf q})=\delta_{ij}-q_iq_j/{\bf q}^2$, and $\gamma$
and $\mu$ are some constants. This ensemble is closely related to that used
by Bausch, Schmitz and Turski \cite{Bausch} in their analysis of the
classical particle diffusion in crystals with randomly distributed
topological defects. It is also identical to the one proposed by
Aronov, Mirlin and W\"{o}lfle \cite{Aronov}. In their paper
$\gamma$ was proportional to the fluctuations of the magnetic
field and $\mu$ was introduced at the intermediate steps of
calculations as a "soft cut-off" in order to avoid the above
mentioned infrared divergencies in the perturbation theory. They
didn't discuss the meaning of this parameter, because
it drops out from their final formula for the diffusion coefficient.
This is true for the calculations  in  the lowest order
in $\gamma$, as we shall see the higher
order terms contain $\mu$ in the very nontrivial way.

One can attach a clear physical meaning to the parameters $\gamma$
and $\mu$ in Eq.~(\ref{correlation}) by analyzing an analogy
between the above discussed model and another model of randomness in which
these coefficients appear in a quite natural way. This auxiliary
model has a well defined physical interpretation and long and
short wavelength  behaviors of both these models are identical.

 Imagine two identical, infinitely thin solenoids (Aharonov -
Bohm flux tubes) piercing our plane at points separated by a
distance  $\ell$. The magnetic flux in one solenoid is
$+\Phi$ and in the second $-\Phi$.
For sake of definiteness we shall call such an object a flux
dipole. On the two-dimensional plane the dipole is characterized by
its position, i. e. the coordinates of its ``center of mass'' and the
angle $\varphi$ between the vector joining the flux tubes and, say,
the $x$ axis.

Now we assume that we have many such dipoles penetrating the plane
and let the  average dipole density be $n$. The position of the $i$'th
dipole ${\bf R}_i$ may be arbitrary and the angle $\varphi_i$ is
homogeneously distributed in the interval $(0,2\pi)$. Notice that by
 construction our disorder ensemble allows only for configurations in
which the total magnetic flux through the plane equals zero. For
such kind of the magnetic disorder we obtain the following
expression for the second moments of the Fourier space components of the
vector potential field $A_i({\bf q})$:
\begin{equation}
 \label{correl1}
 \langle A_i({\bf  q})A_j({\bf q}')\rangle =C({\bf q},{\bf q}')=
(2\pi)^2\delta({\bf q}+{\bf q}') 2n
\left(\frac{\Phi}{\Phi_0}\right)^2 g(|{\bf q}|)
\delta^T_{ij}({\bf q})
\end{equation}
 where
$ g(|{\bf q}|)=(1-J_0(|{\bf q}|\ell))/{\bf q}^2$, $J_0(z)$
is the zero order  Bessel function and $\Phi_0 = e/\hbar c$. The
ensemble average denotes here
\begin{equation}
\langle {\cal O} \rangle = \prod_{i=1}^{N}\left(\frac{1}{2\pi
S}\int_{0}^{2\pi}d\varphi_i \int d^2R_i\,\, {\cal O}\right)
\end{equation}
 where $N$ s the number of dipoles and $S$ is the area of the system
($n=N/S$).

We shall also assume that the magnetic flux $\Phi \ll \Phi_0$, thus we may
neglect higher
order cumulants which are of the order $n\Phi^r$ with ($r\geq4$).
Alternatively we may assume that the fluxes in different dipoles
are mutually independent, gaussially distributed quantities with
zero mean and the second moment equal to $\Phi^2$. In this case
the higher order cumulants also vanish.

   Now discuss some properties of the introduced disorder. If the
condition $\sqrt{n}\ell\gg 1$ is satisfied then on the length scales
shorter than $\ell$ (we call $\ell$ the screening length in the
following)  the magnetic field is spatially uncorrelated
because the magnetic fluxes in randomly placed solenoids are mutually
independent. This follows also from the behavior of $C({\bf q},{\bf q}')$
for
$q\ell \gg 1$. In this limit $g(|{\bf q}|) \rightarrow 1/{\bf q}^2$, as
for the
delta
correlated magnetic field. In the opposite limit, $|{\bf q}|\ell
\rightarrow 0$
, the function $g(|{\bf q}|) \rightarrow const$ and we have the case of
spatially uncorrelated vector potential field.
Comparing the short
and long wavelength behavior of the models defined by
Eq.~(\ref{correlation}) and Eq.~(\ref{correl1}) we see that they are
indeed identical provided $\gamma=2n(\Phi/\Phi_0)^2$ and
$\mu=2/\ell$. Whichever of the limiting behavior of
the vector potential correlation is pertinent to  an actual
physical system depends on the
value of the Fermi wavelength for the electron gas.  In this paper we are
interested in situation when the Fermi wavelength $\lambda_F=2\pi/k_F$
is much smaller than $\ell$, i.e. $k_F\ell \gg 1$.  \\

In the  zero temperature and $\omega \rightarrow 0$ limit Kubo
formula for the real part of the conductivity reads:
\begin{equation}
\label{Kubo}
Re\sigma_{xx}(\omega, {\bf k}=0)=\frac{e^2 \hbar^3}{2\pi m^2 S}Re\Biggl\{
{\rm
Tr}\left[\hat{V}_x\hat{G}_R(E_F+\omega)\hat{V}_x\left(\hat{G}_A(E_F)-
\hat{G}_R(E_F)\right)\right]\Biggr\}
\end{equation}
where
$\hat{V}_x=\frac{1}{i}\partial_x-\frac{e}{\hbar c}A_x({\bf r})$ is
proportional to the $x$'th component of the velocity operator and
$\hat{G}_{R/A}(E)$ are retarded and advanced propagators corresponding
to the one particle hamiltonian (\ref{hamiltonian}). In
position representation they have the form:
\begin{equation}
\hat{G}_{R/A}(E;{\bf r},{\bf r}')=\sum_n\frac{\varphi_n({\bf r})
\varphi_n({\bf r}')}{E-\epsilon_n \pm i\eta}
\end{equation}
where $\varphi_n({\bf r})$ and $\epsilon_n$ are eigenfunctions
and corresponding eigenenergies. In the equation (\ref{Kubo})
{\rm Tr} denotes integration over ${\bf r}$ and $S$ is the system's area. \\

In our analysis we apply the standard method of the averaging
over the disorder \cite{disorder,Abrikosov}. Every Green's
function in Eq.~(\ref{Kubo}) is expanded in the power series of
the components of the vector potential which we treat as a
perturbation. In the second step
every term of the resulting series is averaged according to the
assumed properties of the magnetic field disorder. The exact
summation of that series is impossible. What is usually done
is the extraction and summation of certain classes of diagrams which
are believed to give the most important contribution to the
final result. The criterium of "importance" is based on the
behavior (an index of divergence) of
a corresponding analytical expression in the limit $\omega
\rightarrow 0$ \cite{disorder}. In the present case the most
divergent diagrams  may be divided in two classes. The first class
leads to renormalization of the one particle Green's function
(the averaged Green's function). The second class consists of
the ladder diagrams. These two classes when summed up
give the so called Boltzmann value of the conductivity \cite{disorder}, or
equivalently  the Boltzmann diffusion constant,
 if one uses the density-density response
function instead of Eq.~(\ref{Kubo}) and extracts its low
frequency and long wavelength behavior \cite{Aronov,Falko}. \\
This procedure well established and widely used in electrostatic
potential disorder case, however, suffers here, from certain
difficulty. Namely, the self energy in the averaged Green's
function calculated for the disorder ensemble defined by
Eq.~(\ref{correlation}) is infinite in the limit $\mu \equiv
\frac{2}{l} \rightarrow 0$. The averaged Green's function
in the limit $kl>>1$ equals
\begin{equation}
\label{aver}
\langle G^{R/A}(E,{\bf k})\rangle=1/(E-k^2/2m \pm i
\omega_0/2)
\end{equation}
where $\omega _0 =n(\Phi/\Phi_0)^2kl/(2m) $ and even for
very small density of dipoles $n$ or for very small amplitude of
magnetic field fluctuations $\gamma$ the self energy operator becomes
dominant for large $l$. Of course we realize that the self
energy in the averaged Green's function has no direct physical
meaning (due to the fact that the Greens function is not a gauge
invariant object) thus this infinity may be of no importance. Such a point
of view seems to be justified, because in the final expression
for the diffusion constant (a gauge invariant quantity)
calculated in the ladder diagram approximation the limit $\mu
\rightarrow 0$ may be performed and the diffusion coefficient is
finite. \\
On the other hand the methods of calculations of the diffusion
coefficient (for example by solving the Dyson
equation in the ladder diagram approximation \cite{Aronov}) depend
on the assumption that $\omega_0 \ll E_F$. Thus the
question arises whether the discussed difficulties concerning
the breaking up the perturbation theory are spurious and arising on
the intermediate steps of calculations only do not influence the
final results or they are of physical origin and one should be
very careful in dealing with them.\\
We think that the simplest method to answer this question is to
perform the analysis of the diagrams' divergencies similar to that
presented by Kirkpatrick and Dorfman \cite{disorder} for the electrostatic
potential disorder problem, i. e. to analyze the divergencies
with respect to $\omega$ in the given order of the expansion
parameter $\gamma$. It turns out that it is enough to
consider diagrams of the first order to notice some very
interesting features. These diagrams are shown  in Fig.~1.
Here the continuous lines correspond to
the unperturbed Green's functions $G^0(E,{\bf k})=1/(E-k^2/2m \pm
i\eta)$ and the broken line represents the averaging.
The number of diagrams
is much larger comparing to the case of electrostatic potential
fluctuations case. This is due to first: the appearance  of the second power
of the vector potential in the hamiltonian - this term
must be kept in order to conserve the gauge invariance,
second: the appearance of
the gauge invariant derivatives $\hat{V}$ in Eq. (\ref{Kubo}).
The origin of the first five diagrams 1a - 1e is the
expansion of the Green's function in powers of the vector
potential. In particular, diagrams 1d and 1e correspond to the
term $A^2$ in the hamiltonian. The diagrams 1f - 1i comes from
the expansion of the Green's function and from covariant derivatives,
and the last diagram 1j represents the contraction of two $A$'s
in two covariant derivatives in Eq.~(\ref{Kubo}).\\
The first three diagrams are the lowest order diagrams belonging
to the above discussed classes and although each of them
separately is divergent in the limit $l \rightarrow \infty$, the
sum of them has well defined limit and it constitutes the first
order term in the density expansion for the frequency dependent
conductivity. That sum behaves like $\gamma/\omega^2$. The
diagrams 1d and 1e do not give contribution to the real part of
the conductivity. The contributions from the next four diagrams
cancel. The last diagram 1j has an interesting behavior. Its
analytical expression is proportional to the integral
\begin{equation}
 \int_0^{2\pi} d\theta \frac{1}{k^2+q^2-2kq\cos (\theta) + \mu^2}
\end{equation}
where $k=\sqrt{2m(E+\hbar \omega)}/ \hbar$ and $
q=\sqrt{2mE}/ \hbar$. We see that for $\mu \ne 0$ it is finite
for $\omega = 0$. However for $\mu = 0$ it behaves like
$\gamma/\omega$. Such a behavior suggests possible relationship
between $\omega$ and $\mu$ when both of them are small. \\
Before proceeding further let us notice that up to now
all the calculations have been performed in the Coulomb,
symmetric gauge. The same result however is obtained in the
Landau gauge $A_x=0$. In that case all the diagrams originating
from the covariant derivatives vanish and we are left with
first five diagrams 1a-1e. Again the diagrams 1d - 1e do not
give contribution and for $\mu=0$ the
$\gamma/\omega$ divergence obtained previously from diagram 1j
is contained now in the sum of first three diagrams 1a - 1c. Of
course this is nothing special, it should be so because the
expression (\ref{Kubo}) is gauge invariant. \\

The natural generalization of the diagram 1j is presented at
Fig.~2. Contrary to Fig.~1 the solid lines now represent the
dressed Green's functions. It is interesting that assuming the
diffusive behavior of the electron gas we may calculate the sum
of these two diagrams exactly. Namely, without the broken lines
the sum of two diagrams is closely related to the density -
density correlation function, more precisely it is proportional
to the Kubo relaxation function \cite{Forster} which for ${\bf
q} \rightarrow 0$ and $\omega \rightarrow 0$ behaves like
\begin{equation}
\Phi({\bf q},\omega) = \frac{\left(\frac{m}{2\pi \hbar^2} \right)}
{Dq^2 - i\omega}
\end{equation}
That is why the
analytical expression corresponding to the sum of the diagrams
at Fig.~2 may be calculated and the final result reads:
\begin{equation}
\label{delta}
Re\delta \sigma = \frac{e^2 \gamma}{32\pi^3mD\mu^2}
\frac{\ln\left(\frac{D\mu^2}{\omega}\right) +
\frac{\pi}{2} \left(\frac{\omega}{D\mu^2}\right)}{1+
\left(\frac{\omega}{D\mu^2}\right)^2}
\end{equation}
The Eq.~(\ref{delta}) has no limit for $\omega \rightarrow 0$
($Re \delta \sigma \sim -\ln(\omega)$).\\
It is interesting to write the sum of diagrams from Fig.~2 in
the "time representation"
\begin{equation}
\delta \sigma = \frac{e^2 \gamma}{16\pi^3m}
\int_0^{\infty} dt e^{i\omega t} F(t)
\end{equation}
where
\begin{equation}
F(t)=
\int_0^{\infty} dq \frac{q}{q^2+\mu^2} e^{-Dq^2t}
\end{equation}
is proportional to the current - current correlation function.
From the above we see that for small times comparing to the
diffusion time corresponding to the screening length ($D\mu^2t \ll 1$)
$F(t) \sim \ln(D\mu^2t)$ and for asymptotically large times
($D\mu^2t \gg 1$)  $F(t) \sim t^{-1}$.\\

The obtained result is interesting and not entirely clear.
First, in conjunction with the diagrammatic analysis presented
earlier it suggests that the screening length or the "soft
cut-off" in the terminology of Aronov, Mirlin and W\"{o}lfle
has important physical meaning determining time
characteristics for the system.
The power law  decay of the function F(t) is of the great
interest. 
The $t^{-1}$ behavior suggests that the long time tails are of
quantum origin \cite{disorder}, for the classical Lorentz gas model
would give $t^{-2}$ behavior. To fully appreciate whether this
decay is 
indeed of a quantum mechanical origin requires a detailed
analysis of the velocity - velocity
correlation function for the classical particle in random magnetic
field. To the best of  our best knowledge this is not available at
the moment.\\

We should now return briefly to the discussion of the
applicability of the perturbation theory.
According to Eq.~(\ref{aver})
perturbation theory is applicable if $\hbar \omega_0 \ll E_F
\equiv \gamma l \lambda_F \ll 1$.
Consider now a part of the system, linear dimension of which is
$R \sim \gamma^{-1/2}$ ($R \ll l$).
The typical magnetic field flux contained in that part of the  system
$\Phi \sim \Phi_0$, consequently the typical magnetic field $B \sim
\Phi_0/R^2 \sim \gamma \Phi_0$.
For a particle of energy $E_F$ the corresponding cyclotron
radius $r_c= \frac{c\sqrt{2mE_F}}{eB}\sim
\frac{\sqrt{2mE_F}}{\hbar \gamma} \sim 1/\gamma \lambda_F$.
Thus the condition $\gamma l \lambda_F \ll 1$ may be rewritten
as $r_c \gg l$ what means that the cyclotron radius
corresponding to the local magnetic field fluctuation should be
much larger then the screening length. \\

In conclusion we have studied the behavior of the two
dimensional degenerate electron gas in the presence of the
perpendicular random magnetic field. Analyzing the lowest order
diagrams we have pointed out the importance of the proper choice
of the magnetic field disorder ensemble. Secondly, taking into
account certain class of diagrams we have obtained the long time
tails in the current - current correlation function.

\ack
This work was supported in part by the KBN-grant 2~P03B~126~10 and
the  European Community grant STD-2EC ERBCIPDCT940011.
%


\vskip0.3cm
\noindent{\bf FIGURE CAPTIONS}

\noindent FIG.\ 1.\ Diagrams for the first order perturbation
theory for the conductivity. The continuous lines correspond to
the nonperturbed Green's functions, the broken lines to the
averaging over the disorder, respectively.

\noindent FIG.\ 2.\ Diagrams leading to the nonanalytical
corrections to the conductivity. Here the solid lines represent
the dressed (averaged) Green's functions.

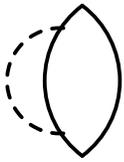
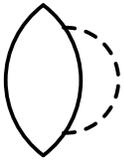
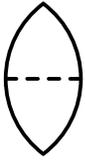
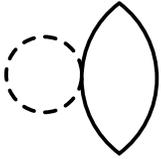
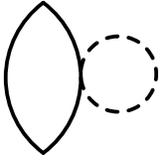
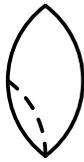
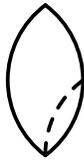
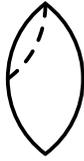
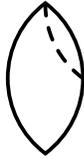
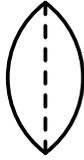
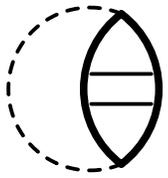
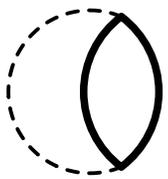
\end{document}